\documentclass[review]{elsarticle}

\usepackage{lineno,hyperref}
\journal{Journal of \LaTeX\ Templates} 

\usepackage[utf8]{inputenc} 
\usepackage[T1]{fontenc}
\usepackage{url}
\usepackage{ifthen}
\usepackage{amssymb}
\usepackage{epsfig}
\usepackage{amsfonts}
\usepackage{pict2e}
\usepackage{color}
\usepackage{amsmath}
\usepackage{graphicx}
\usepackage{mathtools, amsthm, amsfonts, amssymb}
\usepackage{commath}

\usepackage{accents}

\newcommand{\regularize}[1]{\underaccent{\wtilde}{#1}}

\DeclareMathAccent{\wtilde}{\mathord}{largesymbols}{"65}
\DeclareMathOperator{\asymprank}{\underaccent{\wtilde}{R}}
\DeclareMathOperator{\asympsubrank}{\underaccent{\wtilde}{Q}}

\renewcommand{\H}{\mathcal{H}}

\newcommand{\QQ}{\mathbb{Q}}
\newcommand{\NN}{\mathbb{N}}

\newcommand{\RR}{\mathbb{R}}

\newcommand{\E}{\mathbb{E}}

\DeclareMathOperator{\Hom}{Hom}

\newcommand{\jeroen}[1]{#1}

\let\leqx\leqslant

\newcommand{\asympleqx}{%
	\mathrel{%
		\vphantom{\leqslant}%
		\smash{\vcenter{\doasympleqx}}%
	}%
}
\newcommand{\doasympleqx}{%
	\hbox{\ooalign{%
			\noalign{\kern.25ex}
			$\leqslant$\cr
			\noalign{\kern1.25ex}
			\smash{$\sim$}\cr
	}}%
}

\newcommand{\asympdomleq}{%
	\mathrel{%
		\vphantom{\preccurlyeq}%
		\smash{\vcenter{\doasympdomleq}}%
	}%
}
\newcommand{\doasympdomleq}{%
	\hbox{\ooalign{%
			\noalign{\kern.25ex}
			$\preccurlyeq$\cr
			\noalign{\kern1.25ex}
			\smash{$\sim$}\cr
	}}%
}

\newcommand{\asympasympdomleq}{%
	\mathrel{%
		\vphantom{\preccurlyeq}%
		\smash{\vcenter{\doasympasympdomleq}}%
	}%
}
\newcommand{\doasympasympdomleq}{%
	\hbox{\ooalign{%
			\noalign{\kern.25ex}
			$\preccurlyeq$\cr
			\noalign{\kern1.25ex}
			\smash{$\sim$}\cr
			\noalign{\kern0.5ex}
			\smash{$\sim$}\cr
	}}%
}

\newcommand{\domleq}{\preccurlyeq}

\newcommand{\semiring}[1]{#1}

\newcommand{\tand}{\textnormal{ and }}
\newcommand{\tor}{\textnormal{ or }}

\newcommand{\graphs}{\mathcal{G}}

\newcommand{\aspec}{\mathbf{X}}

\let\leqx\leqslant

\newcommand{\doasympgeqx}{%
	\hbox{\ooalign{%
			\noalign{\kern.25ex}
			$\geqslant$\cr
			\noalign{\kern1.25ex}
			\smash{$\sim$}\cr
	}}%
}

\newcommand{\doasympasympleqx}{%
	\hbox{\ooalign{%
			\noalign{\kern.25ex}
			$\leqx$\cr
			\noalign{\kern1.25ex}
			\smash{$\sim$}\cr
			\noalign{\kern0.5ex}
			\smash{$\sim$}\cr
	}}%
}

\newtheorem{Theorem}{Theorem}
\newtheorem{Definition}[Theorem]{Definition}
\newtheorem{Lemma}[Theorem]{Lemma}
\newtheorem{Corollary}[Theorem]{Corollary}

\newtheorem{Remark}[Theorem]{Remark}
\newtheorem{Observation}[Theorem]{Observation}

\def\C{{\mathcal C}}
\def\D{{\mathcal D}}
\def\E{{\mathcal E}}
\def\F{{\mathcal F}}
\def\G{{\mathcal G}}

\def\I{{\mathcal I}}

\def\V{{\mathcal V}}
\def\W{{\mathcal W}}
\def\X{{\mathcal X}}
\def\Y{{\mathcal Y}}

\def\NN{{\mathbb N_0}}
\def\QQ{{\mathbb Q}}
\def\RR{{\mathbb R}}

\vfuzz \hfuzz
\begin{document}
\begin{frontmatter}

\title{Computability of the Zero-Error Capacity with Kolmogorov Oracle}

\author{Holger Boche}
\address{Theoretical Information Technology\\ 
	                  Technical University of Munich\\
                    D-80333 Munich, Germany\\
                    Email: boche@tum.de}
										
\author{Christian Deppe}
\address{Institute for Communications Engineering\\ 
	Technical University of Munich\\
	D-80333 Munich, Germany\\
	Email: christian.deppe@tum.de}
																				


\begin{abstract}

    The zero-error capacity of a discrete classical channel was first defined by Shannon as the least upper bound of rates 
for which one transmits information with zero probability of error.
The problem of finding the zero-error capacity $C_0$, which assigns a capacity 
to each channel as a function, was reformulated in terms of graph theory as a function $\Theta$,
which assigns a value to each  graph. 
This paper studies the computability of the zero-error capacity. 
For the computability, the concept of a Turing machine 
and a Kolmogorov oracle is used. It is unknown if the zero-error capacity is computable in general.
We show that in general the zero-error capacity is semi-computable with
the help of a Kolmogorov Oracle. Furthermore, we show that $C_0$ and $\Theta$ are computable functions if and only if there is a computable sequence of computable functions of upper bounds, 
i.e. the converse exist in the sense of information theory, 
which pointwise converge to $C_0$ or $\Theta$.
Finally, we examine Zuiddam's characterization of $C_0$ and $\Theta$ in terms of Turing computability.
\end{abstract}

\begin{keyword}
zero-error capacity, Kolmogorov Oracle, computability
\end{keyword}

\end{frontmatter}

\section{Introduction}
The zero-error capacity of a discrete classical channel was
introduced by Shannon in \cite{MR0089131}
as the least upper bound of rates 
for which one transmits information with zero probability of error. Investigation of the zero-error capacity of discrete memoryless channels (DMCs) has a long tradition in information theory. Shannon already reduced the problem of determining the zero-error capacity of DMCs to a theoretical graph problem. However, it is generally unclear how Shannon's characterization can be used to compute the zero-error capacity. The zero-error capacity is not known except for special cases even for  graphs. In this work we investigate the Turing computability of the zero-error capacity.

In general, it is not even known if \(\Theta(G)\) is a computable number for every graph \(G\), which is a strictly weaker assertion than the existence of an algorithm that computes \(\Theta(G)\) in dependence of \(G\). 

We consider the Turing computability of the zero-error capacity. 
The concept of a Turing machine provides fundamental performance  limits for today’s digital computers. 
Turing machines have no limitations on computational complexity, have unlimited computing capacity and storage, and 
execute programs completely error-free. Here there are no a priori runtime constraints. Turing machines are the ideal concept to decide whether or not a function (here the zero-error capacity) is effectively computable.
Surprisingly, in information theory the question of Turing computability has attracted very little attention in the past. So far, approaches to bring Turing computability and information theory together have existed in algorithmic information theory (see \cite{AIT}).
For the concept of the Turing machine see \cite{T36, T37, W00}.

 As described above, Shannon reduced the problem of determining the zero-error capacity of DMCs to a theoretical graph problem. 
It turns out, that one has to find the size of a maximum independent
set of a family of graphs to compute the zero-error capacity.
Most  important  results  concerning  zero-error  information  theory  can  be found in the survey paper \cite{KO98}.
Since finding the cardinality of a maximum independent set is a difficult problem, 
we are unlikely to find an efficient algorithm for finding the independence number.
The problem of finding the independence number is an NP-hard optimization problem (see \cite{GJ79}). However, this does not say anything about the computability of this number. 
We show that the zero-error capacity is semi-computable if we allow the Kolmogorov oracle. 
In Section~\ref{Definitions} we introduce an enumeration of  graphs and
give further basic definitions and notations of graph theory and computation theory.
In Section~\ref{Turing} we introduce a Kolmogorov Oracle and show that with this oracle we can compute the zero-error capacity up to any given accuracy. This assumes that we just have an oracle that computes Kolmogorov complexity for free.
An overview of Turing computability with oracles can be found in \cite{H87}.
We  characterize the Shannon's zero-error capacity in Section~\ref{Characterization}
and state some results of \cite{Z19}, discuss the Strassen preorder following \cite{strassen1988asymptotic,Z19}, and discuss the asymptotic spectrum of graphs.
Finally, we show the semi-decidability of the binary relation <~ on graphs with an oracle.
In \cite{Z19} it is shown that 
$\Theta(G)$ equals the asymptotic subrank of the semiring with the
Strassen preorder $\asympleqx$.


\section{Basic Definitions and Concepts}\label{Definitions}

In this section we give basic definitions, notations and concepts. 
We denote the set of natural numbers including $0$ with $\NN$. We need some basic definitions from graph theory. 
\begin{Definition}
	A graph $G$ is a pair $G= (\V,\E)$, where $\V = \{1,2,\ldots,n\} = [n]$ is a set of \(n\) vertices. $\E$ is a set of unordered pairs $\{u,v\}$ of vertices $u,v\in \V$ with \(u \neq v\). 
	The elements of $\E$ are called edges. We write $\V(G)$ for the set of vertices and $\E(G)$ for the set of edges of a 
	graph $G$ . $|G|=|\V(G)|$ denotes the number of vertices and $e(G) =|E(G)|$ denotes the number of edges.
\end{Definition}
We need a computable enumeration of graphs, which we explain later. For this we need the following definitions.
\begin{Definition}
We denote the set of all graphs with $n$ vertices by $\G(n)$ and by $\G=\bigcup_{n=0}^{\infty} \G(n)$ the set of all
graphs.  
\end{Definition}
\begin{Definition}
	The vertices $u,v$ are called adjacent in $G=(\V,\E)$ if $\{u,v\}\in \E(G)$, otherwise $u,v$ are called nonadjacent. An edge $e\in \E(G)$ is incident to a vertex $v\in \V(G)$ if
	$v\in e$. The edges $e,f$ are incident if $e\cap f\not = \emptyset$. If $\{u,v\}\in \E$ then $v$ is a neighbor of $u$.
\end{Definition}
\begin{Definition}
	Let $G= (\V,\E)$ be a  graph with $\V= [n]$. The adjacency matrix $A:=A(G)$ is the $n\times n$ symmetric matrix defined by 
	\[ 
	a_{ij}=\left\{ \begin{array}{ll} 1 & if\ \ \{i,j\}\in \E,\\
	0 & otherwise.\end{array}\right.
	\]
\end{Definition}

We now define an enumeration of the  graphs. We use the 
adjacency matrix. First consider all  graphs with $n$ vertices.
The graph can be described by the values of the adjacency matrix $A$
by knowing $a_{ij}$ with $i<j\in[n]$, because $A$ is symmetric and the 
diagonal has only the entries $0$. Therefore we can represent all
elements of $\G(n)$ by the binary vector
\(
a^{\frac {n^2-n}2}=(a_{12},a_{13},\dots, a_{1n}, a_{23}, a_{24},\dots ,a_{2n},\dots, a_{n-1n})\in \{0,1\}^{\frac {n^2-n}2}.
\)
\begin{Definition}
All $G\in \G(n)$ can be represented by a number between $0$ and \(2^{\frac{n^2 - n}{2}} - 1\) by using the decimal representation of the binary number \(a^{\frac {n^2-n}2}\).
We map the empty graph ($n=0$) to $0$, $G\in \G(1)$ to $1$,\dots,
all $G\in \G(n)$ to the numbers from $\sum_{j=0}^{n-1} 2^{\frac {j^2-j}2}$
to $\left(\sum_{j=0}^{n} 2^{\frac {j^2-j}2}\right)-1$ by keeping the order
as described below. We call this function $\eta:\NN\to \G$ the numbering
of  graphs. 
\end{Definition}
A function from $\NN$ to $\G$ is called partial recursive if it can be computed by a Turing machine, that is, if there exists a Turing machine that accepts input $x$ exactly when $f(x)$ is defined, in which case it leaves the string $f(x)$ on its tape upon acceptance.

It is obvious that $\eta$ is a partial recursive function. 
We can define an order on $\G$ by using
$\eta$: let $G,G'\in\G$, then we set $G\leq_\eta G'$ iff $\eta^{-1}(G)\leq \eta^{-1}(G')$. Ordering relations defined in this manner are of importance for our considerations later on.

Of fundamental importance in describing and studying the zero-error capacity of DMCs are the binary operations, strong product of graphs, and disjoint union of graphs, which we introduce next.
\begin{Definition} Let $G = (\V(G),\E(G))$ and $H = (\V(H),\E(H))$ be  graphs. 
We define the strong product $G \boxtimes H$ by
	\begin{align*}
	V(G \boxtimes H) =& \V(G) \times \V(H)\\
	E(G \boxtimes H) =& \bigl\{ \{(g,h), (g'\!, h') \} :  \\
	&\bigl(\{g,g'\} \in \E(G) \tand \{h,h'\} \in \E(H)\bigr) \\
	&\hspace{0em}\tor \bigl(\{g,g'\} \in \E(G) \tand h=h'\bigr) \\ 
	&\hspace{0em}\tor \bigl(g=g' \tand \{h,h'\} \in \E(H) \bigr)\bigr\}.
	\end{align*}
	
	The $n$-strong product of graph $G$ ($G^{\boxtimes n}=G\boxtimes G\boxtimes\dots\boxtimes G$) 
	satisfies \(\V\big(G^{\boxtimes n}\big)=\{(u_1,\dots,u_n) :u_i\in \V\}\) and $(u_1,\dots,u_n)\not = (v_1,\dots,v_n)$, adjacent 
	if and only if $\big(u_i=v_i \tor \{u_i,v_i\}\in\E(G)\big)$ for all $i$.%
\end{Definition}

\begin{Definition}
	Let $G_1 = (\V_1,\E_1)$ and $G_2 = (\V_2,\E_2)$ be  graphs. 
	We define the disjoint union $G_1\sqcup G_2$ as a graph $G$ with the vertex set
	$\V(G)=\V_1\sqcup \V_2$ and the edge set $\E(G) = \E_1 \sqcup \E_2$,
	where $\sqcup$ denotes the disjoint union of sets.
\end{Definition}

Next we need to introduce appropriate functions on graphs.

\begin{Definition} Let $G\in \G$ be a  graph. An independent set in $G$ is a set of pairwise nonadjacent vertices.  
	A maximum independent set in $G$ consists of the maximum number of pairwise nonadjacent vertices and its size is denoted by $\alpha(G)$.
\end{Definition}

Now we define the zero-error capacity. Therefore we need the definition
of a discrete memoryless channel. In the theory of transmission, the receiver must be in a position to
successfully decode all the messages transmitted by the sender.
\begin{Definition}
	A discrete memoryless channel (DMC) 
	is a triple $(\X,\Y,W)$, where $\X$ is a finite input alphabet, 
	$\Y$ is a finite output alphabet, and 
	$W(y|x)$ with $x\in\X$, $y\in\Y$ is a stochastic matrix.
	The probability for a sequence $y^n\in\Y^n$ to be received if 
	$x^n\in\X^n$ was sent is defined by
	$$
	W^n(y^n|x^n)=\prod_{j=1}^n W(y_j|x_j).
	$$
\end{Definition}

Two sequences $x^n$ and $x'^n$ of size $n$ of input variables are distinguishable by a receiver if  the vectors $W^n(\cdot|x^n)$ and $W^n(\cdot|x'^n)$ are orthogonal. That means if $W^n(y^n|x^n)>0$ then $W^n(y^n|x'^n)=0$
and if $W^n(y^n|x'^n)>0$ then $W^n(y^n|x^n)=0$.
We denote by $M(W,n)$ the maximum cardinality of a set of mutually orthogonal vectors among the $W^n(\cdot|x^n)$ with $x^n\in\X^n$.
The corresponding rate of the channel
is defined by $\frac{\log_2 M(W,n)}n$.
There are different ways to define the capacity of a channel. The so-called pessimistic  capacity is defined as $\liminf_{n\to\infty} \frac {\log_2 M(W,n)}n$
and the optimistic capacity is defined as $\limsup_{n\to\infty} \frac {\log_2 M(W,n)}n$. In general, the pessimistic capacity and the optimistic capacity are different. In the case of zero-error capacity, however, the two size quantities are the same.
A discussion about these quantities can be found in \cite{A06}. 
We define
the zero-error capacity as follows.

\begin{Definition}
	The zero-error capacity of $W$ is:
	\begin{equation}\label{9}
	C_0(W) = \liminf_{n\to\infty} \frac {\log_2 M(W,n)}n
	\end{equation}
\end{Definition}
The zero-error capacity can be characterized in graph-theoretic terms as well. Let $W$ be given and $|\X|=q$.
	To get a formula for $C_0(W)$ Shannon \cite{MR0089131} defined a  graph $G(W)$ for coding with $q=|G|$. In this graph two letters/vertices $x$ and $x'$ are connected, if one could be confused
	with the other because of the channel noise (i.e. there does exist a $y$ such that $W(y|x)>0$ and $W(y|x')>0$).
	Therefore, the maximum independent set is the maximum number of 1-letter messages which can be sent without danger of confusion.  
	In other words, the receiver knows whether the received message is correct or not.
		It follows that $\alpha(G)$ is the maximum number of messages which can be sent without danger of confusion.  
		Furthermore, the definition is extended to words of length $n$ by
		$\alpha(G^{\boxtimes n})$. Therefore, we can give the following graph
		theoretic definition of the Shannon capacity.
\begin{Definition} 
	The Shannon capacity of a graph $G\in\G$ is defined by
	\[
	\Theta(G) \coloneqq \sup_{n\in\NN} \alpha(G^{\boxtimes n})^{\frac 1n}
	\]
\end{Definition}
Shannon got the following.
 \begin{Theorem}[Shannon \cite{MR0089131}]\label{Shannon}
 	\[
 	2^{C_0(W)}=\Theta(G)=\lim_{n \to \infty} \alpha(G^{\boxtimes n})^{\frac 1n}.
	\]
 \end{Theorem}
\begin{Remark}
		This limit exists and equals the supremum \(
		\Theta(G)=\sup_{n\in\NN} \alpha(G^{\boxtimes n})^{\frac 1n}\) by Fekete's lemma \cite{F23}.
\end{Remark}


Theorem~\ref{Shannon} gives no information about whether $C_0(W)$ and $\Theta(G)$ are computable real numbers at all. There are, of course, computable, monotonically increasing sequences of rational numbers, which each converge to a finite limit value, but for which the limit values are not computable numbers and therefore the convergence is not effective (see \cite{Spe49}).

We denote with $C_k\in\G$ the graph with the vertex set $\V_k=\{0,1,\dots, k-1\}$ and
the edge set $\E_k=\left\{ \{u,u\oplus_k 1\}\right\}$, where $\oplus_k$ denotes the
addition modulo $k$. $\C_k$ denotes the set of all isomorphic graphs to $C_k$. 
\begin{Remark}\label{properties}
There is a lot of research into Shannon's zero-error capacity.
We now list some properties that are important for our later considerations.
\begin{enumerate}
	\item For the maximum independent set in a graph it holds that (by definition of $\boxtimes$) 
	\(
	\alpha(G^{\boxtimes n})\geq \alpha(G)^n.
	\)
	\item It is obvious that $\Theta(G_1\boxtimes G_2)\geq \Theta(G_1)\Theta(G_2)$. 
	However (see \cite{alon1998shannon, haemers1979some}), there exist  graphs $G_1 = (\V_1,\E_1)$ and $G_2 = (\V_2,\E_2)$ such that \begin{equation}
	\Theta(G_1\boxtimes G_2)>\Theta(G_1)\Theta(G_2)
	\end{equation} 
	\item Let \(G\) be a graph with \(|\V(G)| \leq 5\) such that \(G\) is \emph{not} isomorphic to \(C_5\). Then, by \cite{MR0089131}, it holds that 
	\[
	\Theta(G)=\alpha(G)\ \ \text{(single letter)}.
	\]
	\item For $G\in\C_5$, it holds by \cite{Lovasz1979shannon}:
	\[
	\Theta(G)=\sqrt{\alpha(G^{\boxtimes 2})}=\sqrt{5}\ \ \text{(multi letter)}.
	\]
	\item Let $S=(\{6\},\emptyset)\in\G$ and $G=S\sqcup C_5$. We know 
	by Zuiddam's characterization in \cite{Z19} that 
	$\Theta(G)=1+\sqrt 5$ (see Theorem~\ref{basic_th} and Remark~\ref{basic_rem}), but there
	does not exist a $n\in\NN$ such that
	\[
	\Theta(G)=\sqrt[n]{\alpha(G^{\boxtimes n})}=1+\sqrt{5}.
	\]
	Therefore, the limit is necessary in Theorem~\ref{Shannon}.
\end{enumerate}
\end{Remark}

\begin{Remark}\label{Ahlswede}
	In his paper \cite{A70}, Rudolf Ahlswede wrote:
	\begin{quote}
		One would like to have a ``reasonable'' formula for $C_0$, which
		does not ``depend on an infinite product space.'' Such a formula is unknown. An answer as: for given $d$ there exists $k=k(d)$ such that $N(nk,0)=(N(k,0))^n$ could be considered ``reasonable''.
	\end{quote}
	The expression $N(n,0)$ denotes the maximal $N$ for which a zero-error code for $n$ exists. Thus, the definition of \(N(n,0)\) in \cite{A70} matches our definition of \(M(W,n)\) for a fixed channel \(W\).
	Ahlswede made this comment in 1970, where the result of Lov\'asz was not known. 
	But by that result and point 5) in the remark, it is clear that such a
	result is not possible.
	If Ahlswede's ``resonable formula'' were correct, we would of course have achieved Turing computability immediately.
	Ahlswede's question about a reasonable formula can be interpreted in the weakest form as a question about Turing computablility. 
\end{Remark}

We would like to make statements about the computability of the zero-error capacity. This capacity is generally a real number. 
Therefore, we first define when a real number is computable. For this we need the following two definitions. We use the concepts of recursive functions (see \cite{Go30,Go34,Kle52, Min61}) 
and computable numbers (see \cite{PoRi17,W00}).
\begin{Definition}\label{ber}
A sequence of rational numbers $\{r_n\}_{n\in\NN}$ is called a computable sequence if  there  exist partial recursive  functions $a,b,s:\NN\to\NN$ 
with $b(n)\not = 0$ for all $n\in\NN$ and 
\[
r_n= (-1)^{s(n)}\frac {a(n)}{b(n)},\ \    n\in\NN. 
\]
\end{Definition}
\begin{Definition}\label{compreal}
A real number $x$ is said to be computable if there exists a computable sequence of rational numbers $\{r_n\}_{n\in\NN}$ such that $|x-r_n|<2^{-n}$ 
for all $n\in\NN$. We denote the set of computable real numbers by $\RR_c$.
\end{Definition}

\begin{Remark}\label{remcomp}
	An equivalent definition of the computability of a real number is: 
	$x \in \RR_c$: $\Leftrightarrow$
	In addition to the sequence $\{r_n\}_{n\in\NN}$ from Definition~\ref{ber}, a partial recursive function $\phi:\NN \to \NN$ can be found such that for all $M \in\NN$ and for all $n\geq \phi(M)$, we have 
	\[
	|x-r_n|<\frac 1{2^M}.
	\] This is precisely the definition of a convergent sequence, whereby the speed of convergence must be effectively computable.
\end{Remark}

We examine the zero-error capacity of general discrete memoryless channels for computability. As described above, each channel can be represented by a  graph. We therefore examine the function $\eta$, which was defined in the first chapter, for computability. We first define when a function is called computable.

\begin{Definition}\label{funcomp} A function $f:\G\to\RR_c$ is called computable
	if $f\circ \eta:\NN\to\RR_c$ is a computable function, meaning
	there are three partial recursive functions $a,b,s:\NN^2\to \NN$ with $b(n,m)\not = 0$ for all $n,m\in\NN$
	such that for all $n\in\NN$ holds:
	
	For all $m\in\NN$ we have
	\[
	\left|f\circ \eta(n)-(-1)^{s(n,m)} \frac {a(n,m)}{b(n,m)}\right|<\frac 1{2^m}.
	\]
	\end{Definition}
\begin{Observation}
 The property of $f$ being computable does not depend on the choice of $\eta:\NN\to \G$, if
 $\eta$ is bijective and partial recursive.
\end{Observation}


The problem is that although we have a representation for $\Theta(G)$ 
as the limit of a convergent sequence, we do not have an effective estimate of the rate of convergence as defined by Definition~\ref{compreal}. So if we calculate the first bits of the binary representation of the number $\Theta(G)$ even for a fixed graph G, this is not possible with only the result of Shannon. (The same observation naturally also applies to the first numbers of the decimal representation of the number $\Theta(G)$.) An approach would now be, e.g., for the decimal representation to derive the best possible lower bound for $\Theta(G)$ from the achievability part and to derive a good upper bound for $\Theta(G)$ from an approach for the inverse part. If the two bounds match for the first $L$ decimal places, then we have determined $\Theta(G)$ for the first $L$ decimal places. Today, even for a fixed graph $G$, there is generally no algorithm that produces a computable sequence of computable numbers as a monotonically increasing sequence of lower bounds and a computable sequence of computable numbers as a monotonically decreasing sequence of upper bounds, both converging to $\Theta(G)$. Indeed, it is unclear whether $\Theta(G)\in\RR_c$ always applies. If this does not apply, then of course such an algorithm cannot exist. It should be noted here that in this discussion we do not require the algorithm to depend on the graph $G$, i.e., an individual algorithm can be developed for each graph $G$. This is exactly possible if $\Theta(\G)\in\RR_c$ applies. This discussion is reflected in the level of knowledge about the behavior of the function $\Theta$. It was a huge step forward that Lov\'asz \cite{Lovasz1979shannon} calculated $\Theta(C_5)$ using the Lov\'asz Theta function. 
Since then, $\Theta(C_7)$ has not been determined and it is not known if the Lov\'asz Theta function is sufficient for this, because there 
is no matching lower bound \cite{PS18, haemers1979some}.
The requirement for the computability of a function $f$ according to 
Definition~\ref{funcomp} is of course much stronger than just the requirement for the property $f:\G\to \RR_c$, because in Definition~\ref{funcomp} the algorithm, i.e. the functions $a$ and $b$, depends of course on the graph.

\section{A Turing machine with a Kolmogorov oracle}\label{Turing}
In the following we study if the zero-error capacity is Turing computable 
with the help of a Kolmogorov oracle. 
This assumes that we just have an oracle that computes Kolmogorov complexity for free.
An overview of Turing computability with oracles can be found in \cite{H87}.

This means we find an algorithm
or Turing machine which can compute the zero-error capacity with the help
of a Kolmogorov oracle. 
More specifically, we compute sharp lower and upper bounds for $C_0$ and 
$\Theta$, whereby we can specify the approximation quality as desired.
Now, we introduce a Kolmogorov oracle. For the definition of
a Turing machine we refer to \cite{Soa87, W00}. 
A Turing machine is a mathematical model of an abstract machine that manipulates symbols on a tape 
according to certain given rules.  
It can simulate any given algorithm and therewith provides a very  powerful model of computation. Turing machines have no limitations on computational complexity, unlimited computing capacity and storage, and execute programs completely error-free. First we need the following definition.
\begin{Definition}
	A subset $\G_1\subset \G$ is called semi-decidable if there is a Turing machine $TM_1$ with the state ``stop'', such that
	$TM_1(G)$ stops if and only if $G\in\G_1$. Therefore $TM_1$ has only one 
	stop state and the Turing machine computes forever, if it does not stop.
\end{Definition}
\begin{Definition}
	A subset $\G_1\subset \G$ is called decidable, if $\G_1$ and 
	$\G_1^c$ are semi-decidable.
\end{Definition}
	$\G_1\subset \G$ is decidable if and only if there is a Turing machine $TM_2$
	with the two states $\{0,1\}$ which gives $TM_2(G)=1$ if and only if
	$G\in\G_1$. The Turing machine $TM_2$ stops for every input $G\in\G_1$. Therefore the characteristic function of the set $\G_1$ is
	described by the definition above. We have the following lemma. 
\begin{Lemma}\label{first}
	Let $\lambda\in\RR_c$, $\lambda\geq 0$. Then the set 
	$\G(\lambda):= \left\{G\in\G: \Theta(G)>\lambda\right\}$
	is semi-decidable.
\end{Lemma}
\noindent {\bf Proof:} We have to define a Turing machine $TM_1$ which is defined on the set $\G$ and stops for the input $G\in\G$ 
if and only if $G\in \G(\lambda)$.
Therefore, consider
\[
f_m(G)=\alpha(G^{\boxtimes 2^m})^{\frac 1{2^m}}.
\]
It holds $f_{m+1}(G)\geq f_m(G)$ and
\[
\lim_{m\to\infty} f_m(G)=\Theta(G).
\]
Therefore, $\Theta(G)>\lambda$ holds if and only if there is a $m_0$, such that
$f_{m_0}(G)>\lambda$. 
We have to show that there is a Turing machine $TM_1$ 
that stops if there is an $m_0$ such that $f_{m_0}(G)-\lambda>0$, and if such a $m_0$  does not exist the Turing machine does not stop.
Therefore, we first define a computable sequence of algorithms.
We know that $\lambda$ is computable, therefore there exists a computable sequence of rational numbers $\{r(n)\}_{n\in\NN}$ such that $|\lambda-r(n)|<2^{-n}$ 
for all $n\in\NN$. Furthermore, there  exist  partial recursive  functions $a,b:\NN\to\NN$ 
with $b(n)\not = 0$ for all $n\in\NN$ and 
\[
r(n)=\frac {a(n)}{b(n)},\ \    n\in\NN. 
\]
Without loss of generality, we assume that $r(n)$ is positive and monotonically decreasing.
Let $A_j$ be the algorithm for computing $f_{j}(G)-\lambda$. 
\begin{itemize}
    \item[$j=0$:] We now specify an algorithm $A_0$ that stops for input $G\in\G$ if and only if $f_0(G)-\lambda>0$ holds.
        It holds $(f_0(G)-\lambda)\in\RR_c$. We set
        \[
        r_0(n):= \left(\frac {b(n)f_0(G)-a(n)}{b(n)}\right).
        \]
        $A_0$ computes $r_0(1)$. If $r_0(1)> \frac 12$ then $A_0$ stops; if not then $A_0$ computes $r_0(2)$. If $r_0(2)> \frac 1{2^2}$ then $A_0$ stops, etc. Assuming the algorithm has not stopped in step $l$, $A_0$ computes $r_0(l+1)$. If $r_0(l+1)>\frac 1{2^{l+1}}$ then the algorithm stops. This algorithm continues to run until it stops.
        It holds: The algorithm stops iff $f_0(G)-\lambda>0$, because:\\
        $\Rightarrow:$ If the algorithm stops, we find an $n_0$ such that
        \[\left(\frac {b(n)f_0(G)-a(n_0)}{b(n_0)}\right)>\frac 1{2^{n_0}}.
        \]
        Therefore $f_0(G)-\left( \frac {a(n_0)}{b(n_0)}\right)>\frac 1{2^{n_0}}$. It holds $|\lambda-r(n_0)|<\frac 1{2^{n_0}}$. We have
        \begin{eqnarray*}
        \frac 1{2^{n_0}} <f_0(G)-r(n_0)&=&f_0(G)-\lambda+\lambda-r(n_0)\\
        &<&f_0(G)-\lambda +|\lambda -r(n_0)|\\
        &<& f_0(G)-\lambda+ \frac 1{2^{n_0}}       
        \end{eqnarray*} 
    Therefore, $f_0(g)-\lambda>0$.\\
        $\Leftarrow:$ If $f_0(G)-\lambda>0$ then there is a $n_0$ such that
        \begin{equation}\label{f} 
        (f_0(G)-\lambda)>\frac 2{2^{n_0}}.
        \end{equation} 
        Therefore, 
        \begin{eqnarray}\nonumber
        	f_0(G)-\lambda &=& f_0(G)-r(n_0)+r(n_0)-\lambda\\\nonumber
        	&\leq& f_0(G)-r(n_0)+|r(n_0)-\lambda|\\\label{g}
        	&<& f_0(G)-r(n_0)+\frac 1{2^{n_0}}.
        	\end{eqnarray}
        By (\ref{f}) and (\ref{g}) we get 
        $\frac 1{2^{n_0}}<f_0(G)-r(n_0)$ and the algorithm stops.
    \item[$j=1$:] We now specify an algorithm $A_1$ that stops for input $G\in\G$ if and only if $f_1(G)-\lambda>0$ holds.
    In this case we set
    \[
     r_1(n)= (f_1(G))^2-\left( \frac {a(n)}{b(n)} \right)^2.
    \]
    $A_1$ computes $r_1(1)$. $A_1$ stops if $r_1(1)>(2r(1)+\frac 12)\frac 12$. If $r_1(1)\leq (2r(1)+\frac 12)\frac 12$, $A_1$ computes $r_1(2)$, etc. Assuming the algorithm has not stopped in step $l$, $A_1$ computes $r_1(l+1)$. If $r_1(l+1)>(2r_1(1)+\frac 12)\frac 1{2^{l+1}}$ then the algorithm stops.
        Again, this algorithm continues to run until it stops.
        It holds: The algorithm stops iff $f_1(G)^2-\lambda^2>0$ and therefore if $f_1(G)-\lambda>0$. This is shown in the same way as with algorithm $A_0$.
        Let us analyse
        the general case for $k\in\NN$ and $j=2^k$.
    \item[$j=2^k$:] Let $C(k)=C(k-1)((r(1)+\frac 12)^{2^{k-1}}+r(1)^{2^{k-1}})$, where $C(0)=1$.
    In this case we set
    \[
     r_k(n)= (f_k(G))^{2^k}-\left( \frac {a(n)}{b(n)} \right)^{2^k}.
    \]
    We now specify an algorithm $A_k$ that stops for input $G\in\G$ if and only if $(f_k(G))^{2^k}-\lambda^{2^k}>0$ and therefore $(f_k(G))-\lambda>0$
    holds.
    It holds $(f_k(G)-\lambda)\in\RR_c$.
    $A_k$ computes $r_k(1)$. If $r_k(1)> C(k)\frac 12$ then $A_k$ stops, if not then $A_k$ computes $r_k(2)$. If $r_k(2)> C(k)\frac 1{2^2}$ then $A_k$ stops, etc. Assuming the algorithm has not stopped in step $l$, $A_k$ computes $r_k(l+1)$. If $r_k(l+1)>C(k)\frac 1{2^{l+1}}$ then the algorithm stops. This algorithm continues to run until it stops.
    It holds: The algorithm stops iff $(f_k(G))^{2^k}-\lambda^{2^k}>0$, because:\\
    $\Rightarrow:$ If the algorithm stops, we find an $n_k$ such that
    \[
    r_k(n)= (f_k(G))^{2^k}-\left( \frac {a(n)}{b(n)} \right)^{2^k}>C(k)\frac 1{2^{n_k}}.
    \]
    First we show that \[|f_k(G)^{2^k}-\lambda^{2^k}-r_k(n)|<C(k)\frac 1{2^{n}}.\]
    It holds for all $n$: 
    \begin{eqnarray*}
    &&|f_k(G)^{2^k}-\lambda^{2^k}-r_k(n)|\\
    &&=|\lambda^{2^k}-r(n)^{2^k}|\\
    &&\leq |\lambda^{2^{k-1}}+r(n)^{2^{k-1}}||\lambda^{2^{k-1}}-r(n)^{2^{k-1}}|\\
    &&\leq \left( (2r(1)+\frac 12)^{2^{k-1}} + r(n)^{2^{k-1}}\right) C(k-1) |\lambda-r(n)|\\
    &&= C(k) \frac 1{2^{n}}.   
\end{eqnarray*}
    The sequence of inequalities follows by induction.
    Therefore, we have
    \begin{eqnarray*}
    	C(k) \frac 1{2^{n_k}} <f_k(G)^{2^k}-r_k(n_k)&=&f_k(G)^{2^k}-\lambda^{2^k}+\lambda^{2^k}-r_k(n_k)\\
    	&<&f_k(G)^{2^k}-\lambda^{2^k} +|\lambda^{2^k} -r_k(n_k)|\\
    	&<& f_k(G)^{2^k}-\lambda^{2^k}+ C(k) \frac 1{2^{n_k}}.       
    \end{eqnarray*} 
    Therefore, $f_k(g)-\lambda>0$.\\
    $\Leftarrow:$ If $f_k(G)-\lambda>0$ then there is a $n_k$ such that
    \begin{equation}\label{D} 
    (f_k(G)^{2^k}-\lambda^{2^k})>C(k)\frac 2{2^{n_k}}.
    \end{equation} 
    Therefore, 
    \begin{eqnarray}\nonumber
    f_k(G)^{2^k}-\lambda^{2^k} &=& f_k(G)^{2^k}-r_k(n_k)+r_k(n_k)-\lambda^{2^k}\\\nonumber
    &\leq& f_k(G)^{2^k}-r_k(n_k)+|r_k(n_k)-\lambda^{2^k}|\\\label{E}
    &<& f_k(G)-r_k(n_k)+C(k)\frac 1{2^{n_k}}.
    \end{eqnarray}
    By (\ref{D}) and (\ref{E}) we get 
    $\frac 1{2^{n_k}}<f_k(G)-r_k(n_k)$ and the algorithm stops.
\end{itemize}
The Turing machine $TM_1$ works with the input $G\in\G$ as follows. First the algorithm $A_0$ is executed. If it stops after the first step, $TM_1$ also stops. If this is not the case, $TM_1$ executes the second step from $A_0$ and in parallel the first step from $A_1$. If one of the two algorithms stops, $TM_1$ also stops. If $TM_1$ has not stopped in the first $k$ steps, then $TM_1$ executes the next steps for the algorithms $A_0, A_1, A_2, \dots, A_{k-1}$ in parallel and starts with the first step of the algorithm $A_k$. $TM_1$ stops when an algorithm $A_l$ stops with $0 \leq l \leq k$. It is clear that $TM_1$ stops iff one of the algorithms $A_l$ stops with $l\in \NN$. 
This applies exactly when $\Theta(G)>\lambda$. \hfill$\blacksquare$\\

The following lemma describes a useful recursive listing of the set $G(\lambda)$.
\begin{Lemma}\label{philemma}
	Let $\lambda\in\RR_c$, $\lambda\geq 0$. There exists a partial recursive function
	$\phi_{\lambda}:\NN\to\G$ with $\phi_{\lambda} (\NN)=\G(\lambda)$ and 
	$\phi_\lambda(k_1)\not= \phi_\lambda(k_2)$ $\forall k_1,k_2\in\NN$, $k_1\not = k_2$, where \(\phi_{\lambda}\) may without loss of generality be assumed to be total on \(\NN\).
	Consequently, the set $G(\lambda)$ is recursively enumerable.
\end{Lemma}
\noindent {\bf Proof:} We first prove the existence of a partial recursive function 
	$\phi_{\lambda}:\NN\to\G$ which satisfies $\phi_{\lambda} (\NN)=\G(\lambda)$ and 
	$\phi_\lambda(k_1)\not= \phi_\lambda(k_2)$ $\forall k_1,k_2\in\NN$, 
	$k_1\not = k_2$. Let 
\[
f_m(G)=\Theta(G^{\boxtimes 2^m})^{\frac 1{2^m}}
\]
and
$\G(\lambda)= \left\{G\in\G: \Theta(G)>\lambda\right\}$.
Let $\eta$ be the enumeration function of the graphs in $\G$ defined above.
$G_i$ with $i\in\NN$ is the $(i-1)$th graph in $\G$ concerning this enumeration.
Let $\D$ be an empty set.
Now we define the following algorithm on the Turing machine $TM_1$.
To compute $f_{m}(G)>\lambda$ we use the algorithm of Lemma~\ref{first}.
\begin{enumerate}
	\item Compute $f_1(G_1)$, if $f_1(G_1)>\lambda$, then $\phi_{\lambda}(1)=G_1$, otherwise add $1$ to $\D$.
	\item Compute $f_1(G_2)$ and if $1\in \D$ compute $f_2(G_1)$.\\
	If $f_2(G_1)>\lambda$ then $\phi_{\lambda}(1)=G_1$ and remove $1$ from $\D$.\\ 
	If $f_1(G_2)>\lambda$ then $\phi_{\lambda}(2)=G_2$, otherwise add $2$ to $\D$.
	\item[$k$] Add $k$ to $\D$ and for all $j\in\D$: 
	\begin{enumerate}
		\item Compute $f_{k-j+1}(G_j)$.
		\item If $f_{k-j+1}(G_j)>\lambda$ then $\phi_{\lambda}(j)=G_j$ and remove $j$ from $\D$.
	\end{enumerate}
\end{enumerate}
It remains to be shown that \(\phi_\lambda\) may, without loss of generality, be assumend to be total on \(\NN\). Denote \(\mathrm{dom}(\phi_\lambda)\) the subset of \(\NN\) for which \(\phi_\lambda\) is defined. There exists a partial recursive bijection \(h : \NN \rightarrow \mathrm{dom}(\phi_\lambda)\), see \cite{So87} for details.
By setting \(\phi_\lambda^{'} := \phi_\lambda\circ h\), we obtain a total partial recursive function 
which satisfies $\phi_{\lambda}^{'} (\NN)=\G(\lambda)$ and $\phi_\lambda^{'}(k_1)\not= \phi_\lambda^{'}(k_2)$ $\forall k_1,k_2\in\NN$, 
	$k_1\not = k_2$.
\hfill$\blacksquare$

The algorithm produces a recursive enumeration of the set $G(\lambda)$.
We need a special enumeration for 
\begin{itemize}
	\item the set $\NN$ and
	\item the set of the partial recursive functions.
\end{itemize}
We start with the set of partial recursive functions from 
$\NN$ to $\NN$.
A listing $\Phi=\{\Phi_i:i\in\NN\}$ 
of the partial recursive functions $f:\NN\to\NN$ is called optimal listing if for any other recursive 
listing $\{g_i:i\in\NN\}$ of the set of partial recursive functions there is a constant $C_1$ such that for all $i\in\NN$ holds: 
There exists $t(i)\in\NN$ with $t(i)\leq C_1 i$  and $\Phi_{t(i)}=g_i$.
This means that all partial recursive functions $f$ have a small Gödel number with respect to the system.
Schnorr \cite{Schnorr} has shown that such an optimal recursive listing of the set of partial recursive functions exists. The same holds true for the set of natural numbers $ \NN $.

\begin{Definition}
For $\NN$ let $u_\NN$ be an optimal listing. For the set $\F$ of partially recursive functions, let $u_\F$ be an optimal listing.
Then we define
\(
C_{u_\F}:\F\to\NN
\) with
\(
C_{u_\F}(f):=\min\{k:u_\F(k)=f\}\) and
\(
C_{u_\NN}:\NN\to\NN
\) with \(
C_{u_\NN}(n):=\min\{k:u_\NN(k)=n\}
\)
as the Kolmogorov complexity of $f$ and $n$ in terms of the optimal listings
$u_\F$ and $u_\NN$.
\end{Definition}

Furthermore, for the set $\G$ we define
\(
C_{u_\G}(G):=\min\{k:\eta(u_\NN(k))=G\}.
\)
This is the Kolmogorov complexity generated by $u_\NN$ and $\eta$. 
\begin{Definition}
On $\F$, $\NN$ and $\G$ we introduce a new order criterion. We sort the elements of these sets in terms of Kolmogorov complexity:
\[
G_1\leq_K G_2 \Leftrightarrow C_{u_\G}(G_1)\leq  C_{u_\G}(G_2)
\]
For $\F$ and $\NN$ we define this analogously.
\end{Definition}

\begin{Definition}
	The Kolmogorov oracle $O_{K,\G}(\cdot)$ 
	is a function from $\NN$ to the power set of the set of graphs that produces a list 
	\[
	O_{K,\G}(n):=( G_1,\dots, G_l), 
	\]
	such that 
	$C_{u_\G}(G_1) \leq \dots \leq C_{u_\G}(G_l)\leq n$ for each $n\in\NN$.
\end{Definition}
\begin{Remark}
	According to our definition of graphs and the set $\G$ with $\eta$,
	 this is the same as the listing $O_{K,\NN}$ of the natural numbers $k$ 
	 with\linebreak $C_{u_{\NN}}(k)\leq n$.
\end{Remark}
Let $TM$ be a Turing machine. We say that $TM$ can use the oracle $O_{K,\G}$ if, for every $n\in\NN$, on input $n$ the Turing machine gets the list $O_{K,\G}(n)$.
With $TM(O_{K,\G})$ we denote a Turing machine that has access to the Oracle 
$O_{K,\G}$.
We have the following Theorem.
\begin{Theorem}\label{Turing1}
	Let $\lambda\in\RR_c, \lambda>0$, then the set $\G(\lambda)$ is decidable with
	a Turing machine $TM^*(O_{K,\G})$. This means there exists a Turing machine
	$TM^*(O_{K,\G})$, such that the set $\G(\lambda)$ is computable with this
	Turing machine with an oracle.
\end{Theorem}

\noindent {\bf Proof:} 
For the proof, we find a Turing machine $TM^*(O_{K,\G})$
such that 
\[
TM^*(O_{K,\G}):\G\to\{0,1\}
\]
and $TM^*(G,O_{K,\G}) = 1$ if and only if $G\in \G(\lambda)$ is true.
Let $\lambda\in\RR_c$ and $\lambda>0$ be arbitrary. We consider the set $\G(\lambda)$ and the function $\phi_\lambda$ of Lemma~\ref{philemma}. $\phi_\lambda:\NN\to\G(\lambda)$ is
a bijective and partial recursive function. 
Therefore, $\phi_\lambda$ is a total function.
Let $\eta^{-1}(\G(\lambda))=\NN(\lambda)$. $\NN(\lambda)$ is a recursively enumerable set, because $\eta$ is a recursive bijection.
$\Phi_\lambda=\eta^{-1}\circ\phi_\lambda:\NN\to\NN(\lambda)$ is bijective and partial recursive. Consequently, \(\Phi_\lambda^{-1}=\phi_\lambda^{-1}\circ \eta: \NN(\lambda) \rightarrow \NN\) is a partial recursive function from \(\NN\) to \(\NN\).
Let $k(\lambda)=C_{u_\F}(\phi_\lambda^{-1}\circ \eta)$. It is clear that $k(\lambda)\in\NN$.
Let $n\in\NN$ be arbitrary and $\hat{n}=\Phi_\lambda(n)$, that means $\Phi_\lambda^{-1}(\hat{n})=n$.
We have 
\begin{equation} \label{I}
C_{u_\NN}(n)=C_{u_\NN}(\phi_\lambda^{-1}\circ \eta(\hat{n}))\leq C_{u_\F}(\phi_\lambda^{-1}\circ \eta)C_{u_\NN}(\hat{n})\leq k(\lambda)\cdot(\hat{n} + c).
\end{equation}
The last inequality is a property of the Kolmogorov complexity proved in \cite{Ma10, MaMa14}.
We set $TM^*(\cdot,O_{K,\G})$ as follows: Let $G\in\G$ be arbitrary. 
We compute $\eta^{-1}(G)$ and $k(G,\lambda):= k(\lambda)\cdot (\eta^{-1}(G) + c)$,
where $k(\lambda)$ and \(c\) are parameters of our algorithm that have to be found analytically, but do not depend on $G$. The oracle
$O_{K,\G}$ is used for the input $k(G,\lambda)$. It creates the list
\begin{align}
	\label{eq:KolmogorovList}
	O_{K,\NN}(k(G,\lambda)):=\big\{ n : C_{u_\NN}(n) \leq k(G,\lambda)\big\}.
\end{align}
If $\eta^{-1}(G)\in \Phi_\lambda\big(O_{K,\NN}(k(G,\lambda))\big)$ then we set $TM^*(G,O_{K,G})=1$.\\
If $\eta^{-1}(G)\not\in \Phi_\lambda\big(O_{K,\NN}(k(G,\lambda))\big)$ then we set $TM^*(G,O_{K,G})=0$.\\
We now show the following two statements. 
\begin{enumerate}
	\item[(A)] For $G\in\G(\lambda)$ holds $ TM^*(G, O_{K,\G})=1$.
	\item[(B)] For $G\not\in\G(\lambda)$ holds $ TM^*(G, O_{K,\G})=0$.
\end{enumerate}
We first show (A). Let $G\in\G(\lambda)$, and set \(\hat{n}_G := \eta^{-1}(G)\). Then it holds:
There exists exactly one $n_G$ with $\Phi_\lambda(n_G)=\hat{n}_G$.
Together with (\ref{I}) it holds:
\[
C_{u_\NN}(n_G)\leq k(\lambda)\cdot(\hat{n}_G + c) = k(G,\lambda).
\]
Therefore, $n_G\in O_{K,\G}(k(G,\lambda))$ and (A) holds.\newline
Now we show (B). If \(G\notin \G(\lambda)\), then \(\eta^{-1}(G) \notin \Phi_\lambda(\NN) = \NN(\lambda)\) and consequently, \(\eta^{-1}(G)\not\in\Phi_\lambda\big(O_{K,\G}(k(G,\lambda))\big)\), since
\(\Phi_\lambda\big(O_{K,\G}(k(G,\lambda))\big) \subset \NN(\lambda)\). Therefore, \(TM^*(G, O_{K,\G})=0\).

It is possible for each graph $G$ to compute with the oracle $O_{K,\G}$ the list
of all graphs $\hat{G}$ with $\eta^{-1}(\hat{G})\leq \eta(G)$ which fulfill $\Theta(\hat{G})>\lambda$.
We can now immediately prove the following corollary from the proof of Theorem~2.

\begin{Corollary}
  Let $\lambda\in\RR_c$ with $\lambda>0$ being arbitrary. There exists a Turing machine
  $TM_*(\cdot ,O_{K,\G}):\G\to \{\H:\H \subset \G\}$ such that for all $G\in\G$ holds:
\[
  TM_*(G,O_{K,\G})=\big\{ \hat{G} : \eta^{-1}(\hat{G})\leq \eta^{-1}(G)\ \rm{and}\ \hat{G}\in\G(\lambda)  \big\}
\]
\end{Corollary}


\noindent {\bf Proof:} 
Given $G\in\G$ as an input, we execute the following algorithm:
\begin{itemize}
	\item Compute $k(G,\lambda)=k(\lambda)\cdot (\eta^{-1}(G) + c)$.
	\item Create the list $O_{K,\NN}\big(k(G,\lambda)\big)$ by using the oracle $O_{K,\NN}$.
	\item Compute $\phi_\lambda\big(O_{K,\NN}(k(G,\lambda))\big) =: M(\lambda,G)$.
	\item Set \(TM_*(G,O_{K,\G}) := M(\lambda,G)\cap\{\hat{G}: \eta^{-1}(\hat{G}) \leq \eta^{-1}(G)\}\). The operation \(M(\lambda,G)\cap\{\hat{G}: \eta^{-1}(\hat{G}) \leq \eta^{-1}(G)\}\) is computable, since \(M(\lambda,G)\) is a finite set and for all
	\(\hat{G} \in M(\lambda,G)\) the relation \(\eta^{-1}(\hat{G}) \leq \eta^{-1}(G)\) is recursively decidable.
\end{itemize}	
We show that $TM_*(\cdot,O_{K,\G})$ has the required properties. That is,
\begin{itemize}
    \item[(A).] If \(\hat{G}\in TM_*(\cdot,O_{K,\G})\), then \(\hat{G}\in\G(\lambda)\) and \(\eta^{-1}(G) \leq \eta^{-1}(G)\);
    \item[(B).] If \(\hat{G}\in\G(\lambda)\) and \(\eta^{-1}(\hat{G}) \leq \eta^{-1}(G)\), then \(\hat{G}\in TM_*(\cdot,O_{K,\G})\).
\end{itemize}
(A). We have \(M(\lambda,G) = \phi_\lambda\big(O_{K,\NN}(k(G,\lambda))\big) \subset \phi_\lambda(\NN) = \G(\lambda)\). Thus, if \(\hat{G}\in M(\lambda,G)\), then \(\hat{G}\in \G(\lambda)\). If \(\hat{G}\) is contained in the list \(TM_*(\cdot,O_{K,\G})\), it must be an element of both \(M(\lambda,G)\) and \(\{\hat{G}: \eta^{-1}(\hat{G}) \leq \eta^{-1}(G)\}\), therefore satisfying both \(\hat{G}\in\G(\lambda)\) and \(\eta^{-1}(\hat{G}) \leq \eta^{-1}(G)\).\newline
(B). From \eqref{I}, we know that all \(n\in\NN\) satisfy \(C_{u_\NN}(n) \leq k(\lambda)\cdot (\Phi_\lambda(n) + c)\). If \(\hat{G}\in\G(\lambda)\) and \(\eta^{-1}(\hat{G}) \leq \eta^{-1}(G)\), we have
\begin{align*}
\exists n_{\hat{G}}\in\NN: \eta^{-1}(\hat{G}) = \Phi_\lambda(n_{\hat{G}})  
\end{align*}
(\(\hat{G} = \phi_\lambda(n_{\hat{G}})\), respectively) and furthermore
\begin{align*}
    C_{u_\NN}(n_{\hat{G}}) 
    &\leq k(\lambda)\cdot (\Phi_\lambda(n_{\hat{G}})+c)\\ &= k(\lambda)\cdot (\eta^{-1}(\hat{G}) + c) \\
    &\leq k(\lambda)\cdot (\eta^{-1}(G) + c) 
    \quad = k(G,\lambda).
\end{align*}
Therefore, \(n_{\hat{G}} \in O_{K,\NN}\big(k(G,\lambda)\big)\), and consequently,
\(\hat{G}\in M(\lambda,G)\). Bu assumption, \(\hat{G}\) also satisfies \(\eta^{-1}(\hat{G}) \leq \eta^{-1}(G)\), such that \(\hat{G} \in TM_*(\cdot,O_{K,\G})\).
\hfill$\blacksquare$\\

\begin{Remark}
  \begin{enumerate}
  	\item The Turing machine $TM_*$ with input $G$ can thus be used to decide for all graphs $\hat{G}$ with $\eta^{-1}(\hat{G})\leq \eta^{-1}(G)$ whether $\Theta(\hat{G})>\lambda$ or 
  	 $\Theta(\hat{G})\leq \lambda$ applies. If $\hat{G}$ is in 
  	 the list $TM_*(G,O_{K,\G})$, then $\Theta(\hat{G})\leq \lambda$ applies. If G is not in the list, then $\Theta(\hat{G})> \lambda$ applies.
  	 \item Note that for $G\in\G$ the Kolmogorov oracle creates the list $O_{K,\G}(k(G,\lambda))$, where the Kolmogorov complexity is related to a Strassen operation listing of the set $\G$.
  \end{enumerate}
\end{Remark}
A second consequence of the Theorem~2 is the following corollary.

\begin{Corollary}
	Let $\lambda\in\RR_c$, $\lambda\geq 0$. Then, the set $\{G:\Theta(G)\leq\lambda\}$ is
	semi-decidable for Turning machines with oracle $O_{K,\NN}$, oracle $O_{K,\G}$, respectively. 

\end{Corollary}
\begin{Remark}
	\begin{enumerate}
	\item Noga Alon has asked if the set $\{G:\Theta(G)\leq\lambda\}$ is semi-decidable (see \cite{AL06}).
	We give a positive answer to this question if we can include the oracle. 
		\item We do not know if $ßTheta$ is computable concerning $TM(O_{K,\G})$. 
\end{enumerate}
\end{Remark}

Let $M\in\NN$ be a number with $2^M\geq|G|$. We set 
$I_{0,M}=[0,\frac 1{2^M}]$ and $I_{k,M}=[\frac k{2^M},\frac {k+1}{2^M}]$ for
$k=1,2,\dots, 2^{2M}-1$.
We have the following theorem.
\begin{Theorem}\label{Turing2}
	There exists a Turing machine $TM^{(1)}(\cdot,O_{K,\NN})$ with
	$TM^{(1)}(\cdot,O_{K,\NN}):\G\to \{0,1,\ldots, 2^{2M}-1\}$  such that for
	all $G\in\G$ with \(|G| \leq 2^{M}\) holds
	\[
	TM^{(1)}(G,O_{K,\NN})= r \Leftrightarrow \Theta(G)\in I_{r,M}
	\]
\end{Theorem}


\noindent {\bf Proof:}
Since \(G \in \G \cap \{G : |G| \leq 2^{M}\}\) holds true by assumption, we have \(\Theta(G) \leq 2^{M}\). Therefore, there exists \(k \in \{0,1,\ldots, 2^{2M} - 1\}\) such that 
\(\Theta(G) \in I_{k,M}\). By Theorem~\ref{Turing1}, there exists for all \(k \in \{0,1,\ldots, 2^{2M} - 1\}\) and \(|G| \leq 2^{M}\) a total Turing machine \(TM_{k,M}(\cdot, O_{K,\NN}) : \G \rightarrow \{0,1\}\) that satisfies 
\begin{align*}
    TM_{k,M}(G, O_{K,\NN}) = 1 \quad \Leftrightarrow \quad \Theta(G) > \frac{k}{2^{M}}.
\end{align*}
Thus, for all \(M\in\NN\) there exists a Turing machine \(TM^{(1)}\)
which
\begin{itemize}
    \item simulates \(TM_{k,M}(\cdot, O_{K,\NN})\) for all \(k\in \{1,\ldots,2^{2M}-1\}\) and creates the list \(\I := \big\{ k: ~1 \leq k < 2^{M}, ~TM_{k,M}(G, O_{K,\NN}) = 1\big\}\);
    \item yields \(TM^{(1)}(G,O_{K,\NN}) = \max\big(\{0\}\cup\I\big)\) as an output.
\end{itemize}
Consequently, we have
\begin{align*}
    \frac{TM^{(1)}(G,O_{K,\NN})}{2^{M}} < \Theta(G) \leq \frac{TM^{(1)}(G,O_{K,\NN}) + 1}{2^{M}},
\end{align*}
which is the desired result.
\hfill$\blacksquare$
\begin{Remark}
 The Turing machine \(TM^{(1)}\) requires multiple queries to the oracle \(O_{K,\NN}\), each with different parameters $k(G,\lambda)= k(\lambda)\cdot (\eta^{-1}(G) + c)$ in the sense of \eqref{eq:KolmogorovList}. The values of \(k(\lambda)\) and \(c\) have to be found analytically in order to construct \(TM^{(1)}\).  
\end{Remark}

 This approach does not directly provide the computability of $\Theta$ through $TM^{(1)}$ with Oracle $O_{K,\NN}$. However, we can compute $\Theta$ with any given accuracy.
 \begin{Remark}
It is not clear if the zero-error capacity is a Turing computable function. So one way to approach the problem is to use a certain oracle and to show the Turing computability with this oracle. The Kolmogorov oracle is in the hierarchy of Turing degrees at the lowest non trivial level. The Kolmogorov oracle is also very interesting because of its importance in algorithmic information theory.

We conjecture that the zero-error capacity is not Turing computable. Therefore, it appears to be an interesting problem to find the weakest oracle which allows the computation of \(\Theta(G)\).
\end{Remark}

 \begin{Remark}\label{remark71}
 	We see that in order to prove the computability of $C_0$ and $\Theta$, we need computable converses in the sense of Theorem~\ref{T71}. The recent characterization of Zuiddam \cite{Z19} using the functions from the asymptotic spectrum of graphs is interesting. We examine this approach with regard to computability in the next section.
 \end{Remark}
 
 \begin{Remark}
 Fekete's Lemma \cite{F23} plays an important role in information theory when determining capacities. For example, it was also used to prove Theorem~\ref{Shannon}. On the basis of the Fekete's lemma, the existence of the limit value \eqref{9} could be shown. 
 Let $\{a_n\}_{n\in\NN}$ be a computable sequence of computable numbers with 
 $a(m+n)\geq a(m)+a(n)$ with $m,n\in\NN$. Then the Fekete's Lemma yields that the computable sequence 
 $\{\frac {a(n)}{n}\}_{n\in\NN}$ of computable numbers has a limit $a_*$ and that 
 \[
 a_* = \sup_{n\in\NN} \frac{a(n)}{n}=\lim_{n\to\infty} \frac{a(n)}n.
 \]
 The question now is whether this limit value $a_*$ is also a computable number, that is, whether the proof of Fekete's lemma is effective. This means that an algorithm for computing $a_*$ can be derived from a suitable proof of Fekete's lemma.
 The following example shows that this is not possible: According to Specker \cite{Spe49}, we can find a computable sequence $\{c(n)\}_{n\in\NN}$ of rational numbers with $c(n)\leq c(n+1)$ for $n\in\NN$ 
 and $c(n)<1$, so that this sequence, converges to a number $c_*>0$ and 
 $c_*\not\in\RR_c$ holds. Then the computable sequence $a(n)=nc(n)$ with $n\in\NN$ fulfills the requirements of the Fekete lemma, because it is 
 \(
 a(n+m)=(n+m)c(n+m)=nc(n+m)+mc(n+m)\geq nc(n)+mc(m)\ \ \forall n,m\in\NN,
 \)
and therefore $\lim_{n\to\infty} \frac {a(n)}n=c_*\not\in\RR_c$. 
Although even $\frac{a(n)}n\leq \frac {a(n+1)}{n+1}$ holds.
This result shows that Fekete's lemma cannot be proven constructively, that is, a form of the axiom of choice is necessary for the proof of Fekete's lemma.
Since Fekete's Lemma is not constructive, we don't even know at the moment whether $\Theta(G)\in \RR_c$ always holds for $G\in\G$. 
 \end{Remark}
 
 \section{Characterization of the Shannon Capacity}\label{Characterization}
 Shannon's characterization of the zero-error capacity according to
 Theorem~\ref{Shannon} can be interpreted as an information theoretical characterization of the achievability part. Of course, this can not 
 be interpreted as an effective (i.e. computable) characterization, since no effective estimate of the speed of convergence is known.
 Zuiddam has recently achieved, based on Strassen's work, a very interesting characterization of Shannon's zero-error capacity, which can be interpreted as a characterization by converse, i.e. a sharp upper bound.
 We now examine Zuiddam's representation in terms of its effective computability. Zuiddam's and Strassen's proofs use Zorn's lemma and are therefore not constructive.
 
 We start with an effective converse in the sense that
 we have a computable sequence of computable upper bounds, so that this sequence
 becomes asymptotically sharp.
 We fully characterize if the functions $C_0$ and $\Theta$ can be computed by Turing machines
 under the conditions specified in Theorem~\ref{T71}. We first need to define a computable sequence of computable functions.
 \begin{Definition}
 	A sequence of functions $\{F_n\}_{n\in\NN}$ with $F_n:\X\to \mathbb{R}_{c}$ is computable if the mapping $(i,x)\to F_i(x)$ is computable. 
 \end{Definition}
\begin{Definition}
	A computable sequence of computable functions  $\{F_N\}_{N\in\NN}$ is called computably convergent to $F$, if there exists a partial recursive 
	function $\phi:\NN\times X\to \NN$, such that for all $M\in\NN$ for all $N\geq\phi(M,x)$.
	\[
	\left| F(x)-F_N(x)\right|<\frac 1{2^M}
	\]
	for all $x\in X$ holds.
	\end{Definition}
We remark that in this case $F:X\to \NN$ is also computable.
 \begin{Theorem}\label{T71}
 	The zero-error capacity $\Theta$ and thus the function $C_0$ is Turing computable if and only if there is a computable sequence $\{F_N\}_{N\in\NN}$ of computable functions $F_N: G \to \RR_c$, so that the following conditions apply:
 	\begin{enumerate}
 		\item For all $N\in\NN$ holds $F_N(G)\geq \Theta(G)$ for all $G\in\G$.
 		\item $\lim_{N\to\infty} F_N(G)=\Theta(G)$ for all $G\in\G$.
 	\end{enumerate}
 \end{Theorem}
 
 \noindent{\bf Proof:} ``$\Rightarrow$'' So $\Theta\circ \eta:\NN\to\RR_c$ is a computable function. According to Definition~\ref{funcomp}, there are partial recursive functions $a, b$ with 
 \[
 \left|\Theta(\eta(n))- \frac {a(n,m)}{b(n,m)}\right|<\frac 1{2^m}.
 \]
 So for all $n\in\NN$ and $m\in\NN$ it is always: 
 \[
 \Theta(\eta(n))\leq \frac {a(n,m)}{b(n,m)}+\frac 1{2^m}.
 \]
 Now $\eta^{-1}:\G\to\NN$ is also computable, so
 \[
 F_m(G):=  \frac {a(\eta^{-1}(G),m)}{b(\eta^{-1}(G),m)}+\frac 1{2^m},\ \ G\in\G,\ \ m\in\NN
 \]
 is a computable function. It applies to all $m\in\NN$ $F_m(G)\geq \Theta(G)$ for all $G\in\G$. Furthermore, we have for all $G\in\G$
 \[
 \lim_{m\to\infty} F_m(G)=\Theta(G).
 \]
 ``$\Leftarrow$'' Let for $M\in\NN$
 \[
 \overline F_M(G):=\min_{1\leq N\leq M} F_N(G).
 \]
 $\overline F_M$ is a computable function. We have
 \[
 \overline F_M(G)\geq \overline F_{M+1}(G)\ \ \forall G\in\G.
 \]
 $\{\overline F_M(G)\}_{M\in\NN}$ is a computable sequence of computable functions.
 We continue to take the function $f_M$ for $M\in\NN$ from the proof of Lemma~\ref{first}. $f_M$ is a computable function and $\{f_M\}_{M\in\NN}$ is a computable sequence of computable functions.
 We consider
 \[
 Q_M(G):=\overline F_M(G)-f_M(G)\ \ G\in\G.
 \]
 It holds $Q_M(G)\geq Q_{M+1}(G)$, $M\in\NN$, and it holds for all $G\in\G$
 \[
 \lim_{M\to\infty} Q_M(G)=0.
 \]
 Furthermore, we consider
 \[
 q_M(n)=Q_M\circ \eta(n),\ \ n\in\NN.
 \]
 $\{q_M\}_{M\in \NN}$ is a computable sequence of computable functions.
 Thus the computable sequences $\{a_M\}_{M\in\NN}$, $\{b_M\}_{M\in\NN}$ of partial recursive functions $a_M:\NN^2\to\NN$ with $M\in\NN$ and $b_M:\NN^2\to\NN$ with 
 $b_M(n,m)\not = 0$ exist for all $(n,m,M)\in\NN^3$ with
 \[
 \left| q_M(n)-\frac {a_M(n,m)}{b_M(n,m)}  \right|\leq \frac 1{2^m}
 \]
 for all $n\in\NN$ and $M\in\NN$.
 Because of the $s^m_n$-Theorem (see i.e. \cite{Soa87}) there exist partial recursive functions $a:\NN^3\to \NN$
 and $b:\NN^3\to \NN$ with $a(n,m,M)=a_M(n,m)$ and $b(n,m,M)=b_M(n,m)$ for $(n,m,M)\in\NN^3$. We now consider the following functions for $(n,M)\in\NN^2$:
 \[
 a^*(n,M)=a(n,M,M)\ \ and\ \ b^*(n,M)=b(n,M,M).
 \]
 $a^*$ and $b^*$ are partial recursive functions from $\NN^2$ to $\NN$.
 It holds
 \[
 q_M(n)\leq \frac {a^*_M(n,m)}{b^*_M(n,m)}+\frac 1{2^M} =: W(n,M).
 \]
 We consider the computable functions
 \begin{eqnarray*}
 	\underline W(n,1) &=& W(n,1)\ \ and\ for\ M \geq 2\\
 	\underline W(n,M) &=& \min\{W(n,M);\underline W(n,M-1)\}.
 \end{eqnarray*}
 $\underline W:\NN^2\to \QQ$ is a computable function. It holds for $M\in\NN$:
 \begin{eqnarray*}
 	q_1(n) &\leq& W(n,1) = \underline W(n,1)\\
 	q_2(n) &\leq& q_1(n) \leq \underline W(n,1)
 \end{eqnarray*}
 and 
 \begin{eqnarray*}
 	q_2(n) &\leq& W(n,2),~\text{therefore}\\
 	q_2(n) &\leq&  \underline W(n,2).
 \end{eqnarray*}
 Let us assume that for $M_0$ holds
 \[
 q_{M_0}(n)\leq \underline W(n,M_0),
 \]
 then this holds also for $M_0+1$.
 Therefore, we have for all $M\in\NN$ and $n\in\NN$
 \[
 q_M(n)\leq \underline W(n,M).
 \]
 Furthermore, $\underline W(n,M+1)\leq \underline W(n,M)$ for $n\in\NN$ and $M\in\NN$.
 For each $n\in\NN$, $\{\underline W(n,M)\}_{M\in\NN}$ is a computable sequence of
 recursive numbers, which is monotone decreasing and
 \[
 \lim_{M\to\infty} \underline W(n,M)=0.
 \]
 For each $K\in\NN$, let $\phi_n(K)$ be the smallest natural number, such that
 \[
 \underline W(n,\phi_n(K))<\frac 1{2^K}.
 \]
 $\phi_n:\NN\to\NN$ is a partial recursive function. $\{\phi_n\}_{n\in\NN}$ is a computable sequence of such functions. Therefore, there exists by the $s^m_n$-Theorem a partial recursive function $\phi:\NN^2\to\NN$ with
 \[
 \phi(n,K)=\phi_n(K), \ \ (n,K)\in\NN^2.
 \]
 Therefore, the function
 \[
 \W^*(n,K)=\underline W(n,\phi(n,K)\ \ (n,K)\in\NN^2
 \]
 is a partial recursive function $W^*:\NN^2\to\QQ$. Thus for $K\in\NN$
 \[
 0\leq \overline F_{\phi(n,K)}\circ \eta(n)-f_{\phi(n,K)} \circ \eta(n) < W^*(n,K)< \frac 1{2^K}.
 \]
 Now, for $M>\phi(n,K)$ we have
 \[
 f_M \circ \eta(n)\geq f_{\phi(n,K)} \circ \eta(n),
 \]
 and for arbitrary $M>\phi(n,K)$ it holds
 \begin{equation}\label{eqA}
 0\leq \overline F_{\phi(n,K)}\circ \eta(n)-f_M \circ \eta(n) \leq
 \overline F_{\phi(n,K)}\circ \eta(n)-f_{\phi(n,K)} \circ \eta(n)< \frac 1{2^K}.
 \end{equation}
 The right hand side of (\ref{eqA}) does not depend on $M$, therefore
 \[
 0\leq \overline F_{\phi(n,K)}\circ \eta(n)-\Theta\circ \eta(n)< \frac 1{2^K}.
 \]
 Therefore, $\Theta\circ \eta$ is a computable function.\hfill$\blacksquare$
 

 In applications, attempts are often made to find suitable representations for certain functions $F:\G\to\RR_c$. Since $F(G)\in\RR_c$ generally applies to a computable function 
 $F$ according to Definition~\ref{funcomp} for $G\in\G$, $F(G)$ can only be approximated effectively. 
 For many questions, however, $F(G) \in \QQ$ applies and thus the number $F(G)$ can always be computed in a finite number of steps. 
 
 Let $F$ be a computable function according to Definition~\ref{funcomp} with $F(G) \in \QQ$ for all $G\in\G$. Is there now a computable function $\zeta:\G\to\QQ$ with $F(G) = \zeta(G)$ 
 for all $G\in\G$? 
 
 This is equivalent to the question: Are there two computable functions 
 $a:\G\to\NN$ and $b:\G\to\NN$ with $b(G) \not= 0$ for all $G\in\G$, 
 such that 
 \begin{equation}\label{bruch}
 F(G)=\frac {a(G)}{b(G)} 
 \end{equation}
 holds for all $G\in\G$?
Next we answer this question negatively.
\begin{Theorem}
 \label{thm:noRecursiveFct}
 There exists a computable function $F:\G\to\QQ$ such that there do not exist computable
 functions $a:\G\to\NN$ and $b:\G\to \NN$ with $b(G)\not= 0$ for all $G\in\G$, such that
 for all $G\in\G$ holds:
 \[
 F(G)=\frac {a(G)}{b(G)}.
 \]
\end{Theorem} 
\noindent {\bf Proof:} 
We first construct a computable function $f:\NN\to\QQ$. 
Let $A\subset\NN$ be a recursive enumerable but not recursive set. 
Let $TM_A$ be the Turing machine that accepts exactly the set $A$. That means $TM_A$ 
stops for the input $n\in\NN$ if and only if $n\in A$ applies. 
We now define for $l\in\NN$:
\begin{equation}
    f_l(n):=\left\{ \begin{array}{ll} \frac 1{2^m} & \textnormal{$TM_A$ stops for input $n$ in $m\leq l$ steps}\\
    \frac 1{2^l} & \textnormal{$TM_A$ does not stop for input $n$ after $l$ steps}.\end{array}\right.
\end{equation}
Here we count the basic steps using the Turing machine $TM_A$. The sequence 
$\{f_l\}_{l\in\NN}$ is of course a computable sequence of functions of the form (\ref{bruch}). 
It is easy to see that the sequence effectively converges to a computable function 
$f_*:\NN\to\QQ$. The following applies: 
\[
f_*(n)>0\Longleftrightarrow n\in A.
\]
Assume there are now $a_*:\NN\to\NN$, $b_*:\NN\to\NN$, $b_*(n)\not= 0$ for all $n$ with 
\[
f_*(n)=\frac {a_*(n)}{b_*(n)}.
\]
Then
\[
a_*(n)=0\Longleftrightarrow n\in A^c.
\]
So, there is an algorithm for testing whether $n\in A^c$, i.e. $A$ is a recursive set. Thus, we have created a contradiction which means the theorem is proven for
$f:\NN\to\QQ$. For $F=f\circ \eta^{-1}$ the proof is clear.\hfill$\blacksquare$\\
We immediately get the consequence that the property (\ref{bruch}) for functions 
is not stable for monotone convergence, because it holds for all $n\in\NN$ for $l\in\NN$: 
\[
f_l(n)\geq f_{l+1}(n).
\]
All functions $f_l$, $l\in\NN$ have the form (\ref{bruch}), 
but not the function $f_*$.
\begin{Remark}
So we see: Every computable function $F:\G\to\NN$ can be effectively approximated by computable sequences of functions according to (\ref{bruch}), but there are computable functions $F$ according to the above theorem that cannot be represented exactly by (\ref{bruch}), i.e., we never have convergence of the sequences in Definition~\ref{funcomp} for all $G\in\G$ in finitely many steps.
\end{Remark}
The proof of Theorem \ref{thm:noRecursiveFct} shows that every computable function $f$ can be represented as a limit of computable monotone decreasing sequences of computable functions of the form 
(\ref{bruch}) and as a limit of computable monotone increasing sequences of functions of the form (\ref{bruch}). The limit function $f:\NN\to\QQ$ generally does not have the form 
(\ref{bruch}). The function from Theorem~5 can therefore be approximated effectively as desired. Although $f(n)\in\QQ$ always applies, it can never be computed recursively 
for all $n\in\NN$ in finitely many steps.

Now we discuss the Strassen preorder, introduce the asymptotic spectrum of graphs and state the result of \cite{Z19}. Finally, we show the decidability of the preorder with a Turing machine using an oracle.
 To state the result of \cite{Z19}
 we need further standard notions like graph homomorphism and graph complement.
 \begin{Definition}
 \begin{enumerate}
 \item Let $G$ and $H$ be graphs.
 A graph homomorphism $f : G \to H$ is a map $f : V(G) \to V(H)$ such that for all $u,v \in V(G)$, if~$\{u,v\} \in E(G)$, then $\{f(u), f(v)\} \in E(H)$. In other words, a graph homomorphism maps edges to edges.
 
 \item The complement~$\overline{G}$ of $G$ is defined by $V(\overline{G}) = V(G)$ and $E(\overline{G}) = \bigl\{\{u,v\} : \{u,v\} \not\in E(G), u\neq v\bigr\}$.
 \item We define the relation $\leqx$ on graphs as follows: let $G \leqx H$
 if there is a graph homomorphism~$\overline{G} \to \overline{H}$ from the complement of $G$ to the complement of $H$.
 \end{enumerate}
 \end{Definition}
 Furthermore, we have to discuss the Strassen preorder, following \cite{strassen1988asymptotic}.
 Let~$(S, +, \cdot, 0, 1)$ be a commutative semiring, meaning that~$S$ is a set with a binary addition operation $+$, a binary multiplication operation~$\cdot$, and elements $0,1\in S$, such that for all $a,b,c \in S$:
 \begin{enumerate}
 	\item $+$ is associative: $(a+b)+c = a + (b+c)$
 	\item $+$ is commutative: $a+b = b+a$
 	\item $0+a = a$
 	\item $\cdot$ is associative: $(a\cdot b)\cdot c = a \cdot (b \cdot c)$
 	\item $\cdot$ is commutative: $a\cdot b = b \cdot a$
 	\item $1\cdot a = a$
 	\item $\cdot$ distributes over $+$: $a\cdot (b+c) = (a\cdot b) + (a \cdot c)$
 	\item $0 \cdot a = 0$.
 \end{enumerate}
 For $n \in \NN$ we denote the sum of $n$ ones $1 + \cdots + 1 \in S$ by $n$.

 Let $\leqx$ be a preorder on $S$, i.e.~$\leqx$ is a relation on $S$ such that for all $a,b,c \in S$
 \begin{enumerate}
 	\item $\leqx$ is reflexive: $a\leqx a$
 	\item $\leqx$ is transitive: $a\leqx b$ and $b\leqx c$ implies $a\leqx c$.
 \end{enumerate}
 Notice that antisymmetry in a preorder is not necessary. The following Definitions are taken
 from \cite{Z19}.
 \begin{Definition}\label{strassendef}
 	A preorder $\leqx$ on $S$ is a \emph{Strassen preorder} if
 	\begin{enumerate}
 		\item $\forall n, m \in \NN$\, $n\leq m$ in $\NN$ iff $n \leqx m$ in $S$ \label{strcond1}
 		\item $\forall a,b,c,d \in S$\, if $a\leqx b$ and $c\leqx d$, then $a + c \leqx b + d$ and $ac \leqx bd$\label{strcond2}
 		\item $\forall a,b \in S, b\neq 0$\, $\exists r \in \NN$\, $a \leqx rb$.\label{strcond3}
 	\end{enumerate}
 \end{Definition}
 
 Let $S$ be a commutative semiring %
 and let $\leqx$ be a Strassen preorder on $S$. 
 We use $\leq$ to denote the usual preorder on $\RR$.
 Let~$\RR_{\geq0}$ be the semiring of non-negative real numbers. 
 \begin {Definition}\label{def:fooI}
 Let $\aspec(S, \leqx)$ be the set of $\leqx$-monotone semiring homomorphisms from~$S$ to $\RR_{\geq0}$,
 \[
 \aspec(S) \coloneqq \aspec(S, \leqx) \coloneqq \{ \phi \in \Hom(S, \RR_{\geq 0}) : \forall a,b \in S\,\, a\leqx b \Rightarrow \phi(a) \leq \phi(b) \}.
 \]
 We call $\aspec(S, \leqx)$ the \emph{asymptotic spectrum} of $(S, \leqx)$.
 \end{Definition}
 \jeroen{Note that for every~$\phi \in \aspec(S, \leqx)$ holds $\phi(1) = 1$ and thus $\phi(n) = n$ for all $n \in \NN$.}
 \begin{Definition}
 	For $a,b \in S$, let $a \asympleqx b$ if there is a sequence $(x_N) \in \NN^\NN$ with $x_N^{\smash{1/N}}\to 1$ when $N \to \infty$ such that for all~$N\in\NN$ we have $a^{N} \leqx b^{N}  x_N$. 
 	We call $\asympleqx$ the \emph{asymptotic preorder} induced by~$\leqx$.
 		 \end{Definition}
 	 Fekete's lemma implies that in the definition of~$\asympleqx$ we may equivalently replace the requirement~$x_N^{\smash{1/N}}\to 1$ when $N\to\infty$ by~$\inf_N x_N^{\smash{1/N}} = 1$.
Let $R$ be the Grothendieck ring of $S$.
 The canonical semiring homomorphism $S \to R : a \mapsto [a]$ is, however, not injective in general, which a priori seems an issue. Namely, $[a] = [b]$ if and only if there exists an element $c \in S$ such that $a+c = b+c$.
 
 To see that noninjectivity is not an issue, we use the following lemma. %
 Proving the lemma is routine if done in the suggested order. A proof can be found in %
 \cite[Chapter~2]{phd}. The following results play a crucial role in Zuiddam's characterization (see \cite{phd,Z19}).
 \begin{Lemma}[\cite{phd}]\label{biglem}
 	Let $\domleq$ be a Strassen preorder on a commutative semiring $T$.
 	Let $\asympdomleq$ be the asymptotic preorder induced by $\domleq$ and let $\asympasympdomleq$ be the asymptotic preorder induced by~$\asympdomleq$. 
 	Then the following are true.
 	\begin{enumerate}
 		\item $\asympdomleq$ also is a Strassen preorder on $T$. \label{23}
 		\item For any $a_1, a_2 \in T$, if $a_1 \asympasympdomleq a_2$, then $a_1 \asympdomleq a_2$.\label{asympasymp}
 		\item For any $a_1, a_2, b \in T$ we have $a_1 + b \asympdomleq a_2 + b$ iff $a_1\asympdomleq a_2$. \label{add_reg}
 	\end{enumerate}
 \end{Lemma}

\begin{Lemma}[\cite{Z19}]%
	\label{str_mainth}
	Let $S$ be a commutative semiring and let $\leqx$ be a Strassen preorder on $S$. Then
	\[
	\forall a,b \in S\quad
	a\asympleqx b \,\,\textnormal{ iff }\,\, \forall\phi \in \aspec(S, \leqx)\,\, \phi(a) \leq \phi(b).
	\]
\end{Lemma}

\begin{Remark}
    In Zuiddams original work, Lemma \ref{str_mainth} is a Corollary.
\end{Remark}

 Let~$\boxtimes$ be the strong graph product, let~$\sqcup$ be the disjoint union of graphs, and
let $K_n$ be the complete graph with $n$ vertices, as defined in the introduction.

\begin{Lemma}[\cite{Z19}]\label{lem1}
	The set $\graphs$ with addition $\sqcup$, multiplication $\boxtimes$, additive unit $K_0$ and multiplicative unit $K_1$ is a commutative semiring. %
\end{Lemma}

\begin{Lemma}[\cite{Z19}]\label{lem2}
	The relation \(\leqx\) on \(\G\) is a Strassen preorder. That is: %
	\begin{enumerate}
		\item For $n, m \in \NN$, $\overline{K_n} \leqx \overline{K_m}$ iff $n \leq m$.\label{grpr_i}
		\item If $A \leqx B$ and $C \leqx D$, then $A\sqcup C \leqx B\sqcup D$ and $A\boxtimes C \leqx B\boxtimes D$.\label{welldef}\label{grpr_ii} %
		\item For $A,B \in \graphs$, if $B \neq K_0$, then there is an $r \in \NN$ with $A \leqx \overline{K_r} \boxtimes B$.\label{grpr_iii}
	\end{enumerate}
\end{Lemma}
We denote by $\asympleqx$ the asymptotic preorder on Graphs induced by
$\leqx$.
 
 Recall the definition of the Shannon capacity $\Theta(G) \coloneqq \lim_{N \to \infty} \alpha(G^{\boxtimes N})^{1/N}$. Thus~$\Theta(G)$ equals the asymptotic subrank $\asympsubrank(G)$.
 One analogously defines the asymptotic clique cover number $\regularize{\overline{\chi}}(G) = \lim_{N \to \infty}\overline{\chi}(G^{\boxtimes N})^{1/N}$, which equals the asymptotic rank $\asymprank(G)$.
 It is a nontrivial fact that the parameter $\regularize{\overline{\chi}}(G)$ equals the so-called fractional clique cover number~$\overline{\chi}_f(G)$. %

 Now we can state the main result of Zuiddam. 
 	\begin{Theorem}[\cite{Z19}]\label{basic_th}
 	$\G$ is a collection of graphs which is closed under the disjoint union~$\sqcup$ and the strong graph product~$\boxtimes$, and which contains the graph with a single vertex, $K_1$.
 	Then we have
 	\begin{enumerate}
 		\item $G \asympleqx H$ iff\, $\forall \phi \in \aspec(\semiring{\G}):\phi(G) \leq \phi(H)$\label{asymp_i}
 		\item $\Theta(G) = \min_{\phi \in \aspec(\semiring{\G})} \phi(G)$.\label{asymp_ii}
 	\end{enumerate}
 \end{Theorem}
\begin{Remark}
The Theorem~\ref{basic_th}, especially point 2, is interesting with regard to the discussion in Remark~\ref{remark71}. For example, if all 
$\phi\in \aspec(\G)$ have the property that they can be computed as functions 
$\phi: G\to \RR_c$ and if we find a recursive subset such that the minimization over this subset gives the zero-error capacity, then we could immediately prove that for $G\in \G$, $\Theta(G) \in \RR_c$ always applies, which is still open up to now, as already mentioned.
So far, however, it is both unclear whether $\phi \in X(\G)$ always applies to 
$\phi: G \to \RR_c$, or whether this function can be computed. The proof in 
\cite{Z19} from Theorem~\ref{basic_th} is not constructive. The Zorn lemma is needed. Only a few functions from the asymptotic graph spectrum are also known today, e.g., the Lov\'asz Theta function, the Fractional Haemers bound, and the Fractional orthogonal rank. For these functions (see \cite{Z19}), except for the Lov\'asz Theta function, it is not clear whether they always fulfill $\phi: G \to \RR_c$, because these are defined by the sequence of suitable functions. It is not clear whether effective convergence occurs here even for fixed $G$.
\end{Remark}

\begin{Remark}\label{basic_rem}
It was observed by Zuiddam \cite{Z19} that the characterization 2. from Theorem~\ref{basic_th} leads directly to the following property of the $\Theta$ function. For any two graphs $G_1, G_2$, we have 
$\Theta(G_1 \boxtimes G_2) = \Theta(G_1)\Theta(G_2)$, if and only 
if $\Theta(G_1 \sqcup G_2) = \Theta (G_1) + \Theta (G_2)$ applies. 
Consequently, we have $\Theta(S\sqcup C_5) = 1+ \sqrt 5$. So the answer to Ahlswede's question in Remark~\ref{Ahlswede} is positive for $d \leq  5$, 
but negative for $  d > 5 $. It is interesting that $k(5) = 2$. 
\end{Remark}

 \begin{Remark}
In this remark, we would like to elaborate on the last sentence of Remark~\ref{basic_rem} because of its importance. In Remark~\ref{properties} we have already discussed that we have a single-letter description for graphs with fewer than five nodes ($\Theta(G) = \alpha(G)$). In contrast, we have a two-letter description ($\Theta(G)=\sqrt{\alpha(G^{\boxtimes 2})}$) for exactly five nodes. In relation to $C_0(W)$, this means that for an alphabet (we assume that the input alphabet has the same number of symbols as the output alphabet) with less than five symbols, we only have to examine the performance function with regard to the channel $W(y|x)$ in order to get the optimal rate. In contrast, for an alphabet with five symbols, we have to consider the performance function of $W(y|x)W(y^*|x^*)$. This is an important phenomenon in information theory. Shannon's result in \cite{S48} on the discrete memoryless channel was also successful because he found a simple single-letter description for the capacity. Often one only considers the difference between a single-letter and a regularized infinite-letter formula in information theory. Therefore the example of the zero-error capacity is very interesting. For the rate function that Shannon has chosen for the zero-error problem, the single-letter representation, partially the two-letter representation and for alphabets larger than 6, the multi-letter representation is necessary to determine the optimal rate. Of course there could be another rate function for which the single-letter representation is always optimal. However, this has not yet been found. If the single-letter representation is optimal, the performance function for the presented blocking approach is automatically additive. With the help of this method it was shown, for example, that Marton coding is optimal for the two-receiver Gaussian vector broadcast channel with common message. For this purpose, the additivity of the rate function for all channels was shown in \cite{14}. There are also results showing that additivity is violated. This showed the sub-optimality of super-position coding for certain three receiver broadcast channels (\cite{13,15}) and for the Han-Kobayashi region for the interference channel \cite{16}. It is very important to understand that there is not only a single-letter or a regularized infinite-letter representation, but also something in between. The zero-error capacity is a good example of this.

In the work \cite{BSP20} another example from information theory is given (identification with feedback) in which connections to problems from pure mathematics and computability are shown.
 \end{Remark}
 
 Now we are prepared to prove the decidability of the preorder with an oracle.
 We have \[
 G \asympdomleq H \Leftrightarrow  \forall \phi\in X(\G)\ \ {\rm holds}\ \ 
 \phi(G)\leq \phi(H)
 \]
 $X(\G)$ is a term for an infinite number of functions $\phi$ and it is
 not clear which $\phi\in X(\G)$ is computable.
 Furthermore, $\asympdomleq$ is a binary relation on $\G\times\G$, a set of pairs of graphs.
 Is this binary relation Turing computable? This means there exists a Turing machine $TM$ with
 \[
 TM:\G\times\G\to\{0,1\}\ \ \ {\rm with}\ \ \ TM((G,H))=1 \Leftrightarrow G\asympdomleq H
 \]
 Our goal is to use a powerful oracle, such that with the help of the
 oracle there exists a Turing machine which computes the binary relation $\asympdomleq$.
We need some notations from the theory of recursive functions.
\begin{Definition}
	Let $\phi_k$, $k\in\NN$, the list of partial recursive Functions. $\phi_k$
	is called total, if the domain of $\phi_k$ equals $\NN$.
\end{Definition}
\begin{Definition}
	Let ${Tot}=\{k\in\NN: \phi_k\ {\rm is\ total\ function}\}$. Then
	$O_{Tot}$ is defined as the following oracle.
	In the calculation step $l$, $TM$ asks the oracle 
	if $k\in O_{Tot}$ is satisfied. $TM$ receives in one calculation step 
	the answer yes or no. The Turing machine uses this answer for the next computation, etc. New queries can always be made to the oracle.
	\end{Definition}
 We have the following theorem.
 \begin{Theorem}
 	$\asympdomleq$ is as a binary operation decidable by a Turing machine
 	$TM(\cdot, O_{Tot})$.
 \end{Theorem}
\noindent {\bf Proof:} 
$\asympdomleq$ is a partial order. We construct the Turing machine we are looking for. To do this, we first examine the behavior of the Strassen preorder on the set $\G \times \G$. Therefore, we use the first point of Theorem~\ref{basic_th} of Zuiddam \cite{Z19} which we presented above. Let $(G,H)\in\G\times\G$ be arbitrary.
Then it holds $G\asympdomleq H$ $\Longleftrightarrow$ $\forall$ $\phi\in X(\G)$ is $\phi(G)\leq\phi(H)$.
This is valid iff 
\begin{center}
(A) $\forall$ $\epsilon>0$ $\exists$ $(n,k)\in\NN^2$
with $k\leq \epsilon n$ and $G^{\boxtimes n}\leqx 2^kH^{\boxtimes n}$ holds.
\end{center}
This is true for $(G,H)\in\G\times\G$ iff 
\begin{center}
	(B) $\forall m\in \NN$ $\exists (n,k)\in\NN^2$ with $km\leq n$ and $G^{\boxtimes n}\leqx 2^k H^{\boxtimes n}$.
\end{center}
Next, we prove this property.
(A) $\Longrightarrow$ (B), because this is required for a subset of (A) for (B).

We show now (A) $\Longleftarrow$ (B). We assume that (B) holds. Let $\epsilon>0$ be arbitrarily chosen and $m_\epsilon\in\NN$ be chosen such that $\frac 1{m_\epsilon}\leq \epsilon$. Then $\exists$ 
$n(\epsilon), k(\epsilon)\in\NN^2$ with 
$k(\epsilon)\leq \frac 1{m_\epsilon}n \leq \epsilon n$ and
\[
G^{\boxtimes n(\epsilon)}\leqx 2^{k(\epsilon)}H^{\boxtimes n(\epsilon)}.
\]
Therefore, (A) holds, because $\epsilon$ is chosen arbitrarily.

Now for $(G,H)\in \G\times\G$ and $m,n,k\in \NN^2$, the function
\begin{align}
    F((G,H),m,n,k)= \begin{cases}
                        1 \quad &\text{if}~ km\leq n~\text{and}~ G^{\boxtimes n}\leqx 2^kH^{\boxtimes n} \\
                        0 \quad &\text{otherwise}
                    \end{cases}
\end{align}
is primitive recursive.
Therefore, it holds $G\asympdomleq H$ iff for $(G,H)$ it holds that for all $m\in\NN$ we 
can find $(n,k)\in\NN^2$ such that $F((G,H),m,n,k)=1$ holds.
We now use the $s^m_n$-Theorem (see \cite{Soa87}) of the theory of recursive functions.
It holds: There is a unique recursive function $q:\G\times\G\to \NN$ with
\[
\phi_{q(G,H)}(m)=\Phi((G,H),m),
\]
where
\[
\Phi((G,H),m)=\left\{\begin{array}{ll} 1 & \forall m_1\leq m \exists (n,k): F((G,H),m_1,n,k)=1\\ not\ defined & otherwise
\end{array}\right.
\]
is a partial recursive function. 
Therefore, it holds $G\asympdomleq H$ $\Leftrightarrow$ for all $m$ holds:
$\forall m_1\leq m \exists (n,k):F((G,H),m_1,n,k)=1$. 
This holds iff $\phi_{q(G,H)}$ is a total recursive function.
Therefore, $G\asympdomleq H$ iff $q(G,H)\in Tot$. 
We are now ready to define the Turing machine: Take $(G,H)\in\G\times\G$ to be
arbitrary. Compute the numbers $q(G,H)\in\NN$. Test now if $q(G,H)\in Tot$ or not. Here we used the oracle $Tot$. If $q(G,H)\in Tot$, then $TM((G,H),Tot)=1$.  If $q(G,H)\not\in Tot$, then $TM((G,H),Tot)=0$. $TM((G,H),Tot)=1$ is only valid if
$G\asympdomleq H$, then and only then $q(G,H)\in Tot$ is fulfilled. Therefore, this Turing machine has the behavior required in the theorem.
\hfill$\blacksquare$

\section{Conclusions and Discussions}

It is not clear if the zero-error capacity is a Turing computable function. So one way to approach the problem is to use a certain oracle and to show the Turing computability with this oracle. The Kolomogorov oracle is in the hierarchy of Turing degrees at the lowest non trival level. The Kolmogorov oracle is also very interesting because of its importance in algorithmic information theory.

We even conjecture that the zero-error capacity is not Turing computable.  So that it seems to be an interesting problem to find the weakest oracle.

The  $O_{T_{ot}}$ Oracle is a very strong one. It would be desirable to show the theorem using a weaker oracle.

\section*{Acknowledgments}

Holger Boche was supported by the Deutsche Forschungsgemeinschaft (DFG, German
Research Foundation) under Germany’s Excellence Strategy - EXC
2092 CASA - 390781972. Christian Deppe was supported by the Bundesministerium 
f\"ur Bildung und Forschung (BMBF) through grant 16KIS1005.

We thank Yannik B\"ock for his helpful and insightful comments.

\end{document}